# Interstellar Spectra of Primary Cosmic Ray Nuclei from H through Fe Measured at Voyager and a Comparison with Higher Energy Measurements of These Same Nuclei Made Near the Earth – An Interpretation of the Spectra from ~10 MeV/nuc to Over 100 GeV/nuc Using a Leaky Box Propagation Model

**W.R. Webber**

New Mexico State University, Astronomy Department, Las Cruces, NM 88003, USA

**ABSTRACT**

Utilizing new Voyager measurements at lower energies and higher energy spacecraft measurements near the Earth, the interstellar spectra of primary cosmic ray nuclei from H to Fe have now been determined from ~10 MeV/nuc to > 100 GeV/nuc. These measurements are compared with the predictions from a Leaky Box propagation model. It is found that above ~50-100 MeV/nuc the spectra of all the nuclei, H, He, C, O, Ne, Mg, Si and Fe are well connected between 100 MeV/nuc and 10 GeV/nuc and above by simple source rigidity spectra proportional to $P^{-2.28}$, with the exponent independent of rigidity and using a rigidity dependent diffusion coefficient ~$P^{0.50}$ above ~1.0 GV. This leads to intensities and spectra ~$P^{-2.78}$ at ~100 GeV/nuc and above, which are consistent with new AMS-2 and PAMELA measurements of H, He and C to within $\pm$ 10%.

Below 50-100 MeV the spectra of these primary charges fall into two groups. The spectra of primary nuclei with $Z \geq 6$ fall more rapidly at low energies than those of H and He and more rapidly than would be expected from a pure diffusive LBM, where energy loss by ionization is important at low energies. Increasing the fraction of ionized hydrogen in the interstellar medium from 15% to 30% does not satisfactorily explain the systematics of the intensity decreases of the heavier primary nuclei at low energies. Other processes beyond those assumed in a simple leaky box model, such as a lack of cosmic ray sources near the Sun are being investigated as the source of this more rapid intensity decrease of these nuclei.

The spectra of H and He are very similar to each other but both contain more low energy particles than would be expected from a correspondence with the heavier nuclei spectra. We believe that this may indicate the presence of a local (heliospheric) component of these nuclei in addition to the galactic component.

**Introduction**

The interstellar spectra of the more abundant heavier nuclei H through Fe have now been measured beyond the heliopause (HP) below a few hundred MeV/nuc by the Voyager 1 spacecraft (Cummings, et al., 2015). These spectra and intensities can now be compared with those measured near the Earth at energies above 10 GeV/nuc where the solar modulation is small to obtain interstellar spectra from a few MeV/nuc up to above 100 GeV/nuc, nearly 5 orders of magnitude. This enables a major step to be taken beyond earlier studies such as that by e.g., Maurin, Putze and Derome, 2009, in their study of B/C ratio as a function of energy.

The energy spectra of these primary source nuclei H, He, C, O, Ne, Mg, Si and Fe studied in this paper appear to have two distinct energy regimes:

1) Above ~50-100 MeV/nuc the interstellar intensities of all of these nuclei measured at Voyager connect well with those measured above ~10 GeV/nuc at Earth where solar modulation effects are small (e.g., AMS-2 – Aguilar, et al., 2014, Oliva, et al., 2015; Haino, et al., 2015; PAMELA – Adriani, et al., 2011, 2014; HEAO-2 – Engelmann, et al., 1990; HEAO-3 – Binns, et al., 1988 and others above 100 GeV/nuc, see summaries by Ahn, et al., 2010 and Seo, 2012), in such a way that the overall source spectrum from ~100 MeV/nuc to over 100 GeV/nuc (0.5-200 GV rigidity) for all of these primary nuclei can be represented in a Leaky Box Model (LBM) by a simple rigidity spectrum with a spectral index = -2.28 independent of rigidity, assuming a rigidity dependence of the diffusion coefficient to be ~$P^{0.50}$ above ~1.0 GV (see also Webber and Higbie, 2015a on the H/He ratio). Therefore, at high energies the measured spectra of all of these primary nuclei will have an exponent =-2.78$\pm$ 0.05 as is observed (see above references).

2) Below ~50-100 MeV/nuc the intensities of the primary nuclei with Z $\geq$ 6 decrease more rapidly than would be expected from pure diffusive Leaky Box Models where the main energy loss processes at low energies is by ionization as these nuclei pass through interstellar matter. The H and He nuclei spectra are very similar to each other but contain more lower energy particles relative to the Z $\geq$ 6 nuclei than would be expected in a simple LBM propagation.

We will discuss both of these energy regions in detail in this paper with special emphasis on the lower energy regime. The availability of new data on the spectra of C, O, Mg, Si and Fe down to ~10 MeV/nuc from a 24 month study on V1 after the heliopause crossing (Cummings, et al., 2015) is of essential importance in this study.

## The Data

In Figure 1 we show spectral data for H, He, C, (Ne+Mg and Si) and Fe nuclei from ~10 MeV/nuc to ~1 GeV/nuc. In Figure 2 we show these spectra above 1 GeV/nuc with the intensities x $E^{2.5}$. The data includes only Voyager data below ~1 GeV/nuc. Above 1 GeV/nuc, PAMELA and AMS-2 data are shown as solid colored blue and red lines connecting the data points for H, He and C nuclei (Adriani, et al., 2011, 2014; Aguilar, et al., 2014; Oliva, et al., 2015; Haino, et al., 2015). For C, (Ne+Mg and Si) and Fe nuclei the data from HEAO-2 (Engelmann, et al., 1990) purple line and HEAO-3 (Binns, et al., 1988) orange line, and several other higher energy experiments are shown. These data come from the data summaries from Ahn, et al., 2010 and Seo, 2012, and are shown as solid colored blue and red lines connecting the data points. There is evidence for systematic variations of $\pm$ 10% in the data for some of the heavier nuclei, e.g., the C spectra and also the Fe spectra.

The predictions, shown as black lines, are from a Leaky Box model with source spectra -2.28 for all components. The details of this model have been noted in the introduction and will be discussed more fully in what follows.

For a different perspective on the relative spectra of the heavier nuclei we show in Figure 3 the C/Mg ratio and in Figure 4 the C/Fe ratio as a function of energy. In Figures 3 and 4 we show the galactic propagation calculations (LBM) for a mean material path length $\lambda$=26.5 $\beta$ $P^{-0.50}$ (g/cm$^2$) > 1.0 GV in the galaxy, for source spectra j(P) ~$P^{-2.28}$ and for a diffusion coefficient ~$P^{0.5}$ above 1 GV. The choice of these values will be discussed later.

The propagation calculations in Figure 3 are consistent with the slowly varying measured C/Mg ratio from about 8 at 10 MeV/nuc to ~4.5 near 10 GeV/nuc. This is due to the changing value of the mean path length, $\lambda$, with energy (rigidity). For the C/Fe ratio in Figure 4, these energy dependent changes are more rapid, varying from ~20 at 10 MeV/nuc to ~7 near 100

GeV/nuc. The calculated C/Fe ratio also shows a turn up below 20-30 MeV/nuc. This turn up is due partly to ionization energy loss, considering a medium that is 15% ionized. A doubling of this ionized fraction to 30% will make a slight change in the shape of the C/Mg and C/Fe dependence at low energies as illustrated in the figures.

**Diffusive Cosmic Ray Propagation in the Galaxy**

The galactic propagation model used in this paper is the same LBM model used by Webber and Higbie, 2009, and has evolved from earlier models (Webber and Rockstroh, 1998). The reader is referred to these papers for details. Since the acceleration term in the LBM is set equal to zero, the key parameters affecting the spectral shape and intensities of the propagated particles are:

The mean path length, $\lambda$, in g/cm$^2$ as a function of rigidity (energy/nuc) is determined from a fit to the B/C ratio as shown in Figure 5. This fit includes new measurements of the B/C ratios by Adriani, et al., 2014 and Oliva, et al., 2015, up to several hundred GeV/nuc. It is $\lambda = 26.5 \ \beta R^{-0.5}$ above 1.0 GV up to ~120 GV where the B/C ratio is 0.125 implying that the path length is still 2.5 g/cm$^2$ at this rigidity. Above ~120 GV the path length begins to flatten.

The dependence used below 1.00 GV (~120 MeV/nuc for A/Z = 2.0 particles) is the dependence used to fit the Voyager spectra of H and He nuclei and also the abundances of secondary nuclei, $^2$H, $^3$He and B measured by Voyager at lower energies and is discussed below.

At energies below ~50 MeV/nuc, the loss term, which is mainly the ionization energy loss as the various nuclei traverse the interstellar matter, should become more important.

The uncertainties in $\lambda$ above ~1.0 GV, which mainly depend on fitting the observed cosmic ray B/C ratio as a function of energy, are also described in Webber and Higbie, 2015 a,b. These papers also provide details of the cross sections used in the propagation calculation, including updates, that are necessary for propagation below ~100 MeV/nuc. In this LBM the path length distribution is a simple exponential at each mean path length. The average density is taken to be 0.36 H plus 0.04 He nuclei per cm$^3$. For the H fraction, 15% is ionized.

In our calculations we have used source rigidity spectra ~$P^{-2.28}$ which, along with a diffusion coefficient ~$P^{0.50}$, leads to an exponent of the rigidity spectrum ~-2.78 for all charges at high rigidities. These parameters provide spectra that are excellent fits to the unmodulated H and He spectra at those for other primary nuclei from 50-100 MeV/nuc to ~100 GeV/nuc and above that are shown in Figures 1 and 2 and also to the broad maxima in the differential spectra of H and He nuclei at ~50 MeV/nuc which are observed by Voyager (Stone, et al., 2013). A change in the rigidity dependence of the diffusion coefficient from the high energy dependence ~$P^{0.5}$ to one ~$P^{-0.5}$ (one power in the exponent) at a value of $P_0$= 1.0 GV leads to a path length dependence which is ~beta$^{3/2}$ where beta = v/c at lower rigidities.

The effects of solar modulation at the time of the individual measurements are evident in Figures 1 and 2.

Note also, however, that the data for all of the $Z \geq 6$ nuclei is systematically lower than the LBM predictions below ~50 MeV/nuc in Figure 1.

## Discussion – A Comparison of Differences Between the Propagated and Observed Local Interstellar Spectra for H – Fe Nuclei Below ~100 MeV/nuc

To examine the spectral differences between H, He and the heavier primary nuclei more closely, we show in Figure 6 the ratio of the measured intensities at each energy to the intensities measured at 100 MeV/nuc for each charge. The energy of 100 MeV/nuc acts as a very sensitive normalization point for the shapes of the different spectra at low energies because it is near the maximum in the differential spectra. The differences in the relative low energy spectra of H and He and the spectra of heavier nuclei and above are clearly obvious in Figure 6.

As a result of these spectral differences we have constructed two separate Figures. Figure 7 shows the normalized measured spectra of heavier primary nuclei C, O, (Ne+Mg+Si) and Fe, along with the LBM predictions for $P_0$ = 1.0 GV for these nuclei, also normalized at 100 MeV/nuc. Figure 8 shows only the H and He normalized spectra along with similar LBM predictions normalized at 100 MeV/nuc.

Let us consider Figure 7 with the $Z \geq 6$ spectra first. It is evident that the spectra of these primary nuclei are all deficient in particles below 50 MeV/nuc relative to the LBM predictions for a $P_0 = 1.0$ GV. The data accuracy allows us to specify the difference quite accurately for C and O nuclei. A change of the ionized gas fraction from the standard 15% to 30% produces a small energy and charge dependent change which is much less than is observed.

Consider next Figure 8 showing the H and He nuclei spectra only. In this case the LBM predictions for a $P_0 = 1.0$ are in much closer agreement with the measured spectra normalized at 100 MeV/nuc. In other words, the deficiency relative to a LBM calculation which is evident for $Z \geq 6$ nuclei is not evident for H and He nuclei.

The spectra of H and He themselves are also nearly identical to each other when H to He intensity ratio is taken to be 12.0 at 100 MeV/nuc. This statement is equivalent to the statements in Stone, et al., 2013 and also in Webber and Higbie, 2015a, in their H and He paper, that the H/He ratio is a constant with value = 12.0 below a few hundred MeV/nuc.

A comparison of Figures 7 and 8, along with Figure 1, therefore illustrates the differences in the spectra of low energy H, He nuclei and the heavier $Z \geq 6$ primary nuclei in contrast to their close correspondence at energies above 100 MeV/nuc up to 100 GeV/nuc and above. This is the goal of this paper. The reasons for such a difference in spectra are complex and will be described in follow up papers which examine the various possibilities for this difference. Already we have noted (Webber and Higbie, 2015b) that the difference between the He and C spectra can be best explained by an excess of He nuclei, above that for galactic cosmic rays only, at low energies.

**<u>Summary and Conclusions</u>**

In the time since August, 2012 when V1 crossed into interstellar space, sufficient statistics have now been obtained to study the spectra of primary or "source" nuclei from H to Fe from a few MeV/nuc to a few hundred MeV/nuc. This includes H, He, C, O, Ne, Mg, Si, and Fe nuclei. The correspondence of energy spectra of these nuclei is described in terms of two energy regions; one below 50-100 MeV/nuc and another above this energy. Above ~100 MeV/nuc the unmodulated spectra of all of these nuclei obtained at V1 and also from Earth orbiting spacecraft (including H with A/Z = 1.0) can be described between 100 MeV/nuc and 100 GeV/nuc (~1 to 200 GV) in a Leaky Box propagation model in terms of simple rigidity source spectra $\sim P^{-2.28}$ (with the exponent independent of rigidity). The diffusion coefficient that applies to this propagation is $\sim P^{0.50}$ thus giving observed spectra $\sim P^{-2.78}$ at high rigidities as is generally observed (see Figure 2 and references). The measurements of intensities at V1 at 1 GV ($\sim$120 MeV/nuc and below) and also from spacecraft between $\sim$10 and at least 200 GV where the solar modulation is small, are thus consistent at both ends of this rigidity range within an uncertainty $\sim\pm$ 10% or $\pm 0.04$ in the exponent of the spectra that are required to match the V1 and high rigidity intensities.

At energies below 50-100 MeV/nuc, the experimental situation appears to be more complex. The observed differential spectra of primary nuclei from C to Fe decreases more rapidly than those of H and He at these lower energies and also more rapidly than predicted by a simple LBM. The spectra of H and He nuclei are very similar to each other, but appear to have an excess of low energy particles relative to the spectra of C, O and heavier nuclei.

The H and He differential intensities, in fact, actually continue to increase below $\sim$100 MeV, reaching a peak around 30 MeV/nuc as seen most clearly in Figure 7. At 10 MeV/nuc the intensities of H and He are both still higher than those observed at 100 MeV/nuc. For C-Fe nuclei the intensities at $\sim$10-20 MeV/nuc are only about 0.5 of those observed at 100 MeV/nuc.

Below 50-100 MeV/nuc Leaky Box propagation models with $P_0 \sim$1.0 GV thus "appear" to fit the observed H and He spectra quite well but do not describe well the low intensities observed for the C-Fe primary nuclei. The observed intensities of all of these heavier nuclei are lower than the LBM predictions based on H and He nuclei.

To fully interpret this inconsistency between H and He nuclei spectra and heavier nuclei spectra at lower energies only, modifications to a simple Leaky Box Model seem to be needed. These modifications will be discussed in companion papers with titles as follows:

1) "Evidence for a low energy component of H and He nuclei beyond the Heliopause"; See also, "The cosmic ray helium and carbon nuclei spectra measured by Voyager 1 at low energies and Earth based measurements of the same nuclei up to over 100 GeV/nuc compared to a Leaky Box Propagation Model: Concordance at high energies but a possible excess He component at low energies", http://arXiv.org/abs/1503.05891

2) "A deficiency of short path lengths in the galactic diffusion process for cosmic rays and its relationship to the distribution of the sources of cosmic rays"

These two papers will further enable one to understand the source of the inconsistency in the spectra of H and He nuclei and those of heavier nuclei at energies below ~50-100 MeV/nuc within the framework of a modified LBM model with a deficiency of short path lengths and a possible excess of low energy H and He nuclei above that predicted using a simple LBM.

**Acknowledgements:** The author appreciates discussions with his Voyager colleagues Ed Stone, Alan Cummings, Nand Lal and Bryant Heikkila, on the Voyager data and its implications. This work world have not been possible without the assistance of Tina Villa.

# FIGURE CAPTIONS

**Figure 1:**  Measured spectra for H, He, C, (Ne+Mg+Si) and Fe nuclei from V1 beyond the heliopause (Cummings, et al., 2015).  The black lines are the intensities for H, He, C, (Ne+Mg+Si) and Fe nuclei calculated using a Leaky Box Model with $\lambda = 26.5 \ \beta P^{-0.5}$ above 1.0 GV and a dependence $\lambda = 40.9 \ \beta^{3/2}$ below 1.0 GV, using source spectra of the form $dj/dP \sim P^{-2.28}$ with the exponent independent of rigidity.  The spectrum measured by PAMELA (Bozio, et al, 2014) in 2009 at the time of maximum intensity at Earth is shown as a blue curve.

**Figure 2:**  Data above 1 GeV/nuc is from AMS-2 for H, He and C (Red) (Aguilar, et al., 2015; Owen, et al., 2015; Haino, et al., 2013), PAMELA for H, He and C (blue) (Adriani, et al., 2011, 2014), from HEAO-C for C, O, Mg, Si and Fe (purple) (Engelmann, et al., 1990) and from HEAO-3 for Fe (orange) (Binns, et al., 1988).  Above 100 GeV/nuc additional data is from Ahn, et al., 2010 and Seo, 2012, (purple).  The black shaded regions between 1-10 GeV/nuc represent the effects of solar modulation.  The black lines are the calculated spectra.

**Figure 3:**  Propagation calculations and measurements of the C/Mg ratio.  The propagation calculations are described in the text.  Propagation calculations showing a 15% and 30% fraction of ionized interstellar H are also shown.

**Figure 4:**  Same as Figure 3 except the C/Fe ratio is shown.

**Figure 5:**  The B/C ratio as a function of energy/nuc.  The data points are from AMS-2 (Oliva, et al., 2015).  The solid line is the LBM calculation for $\lambda=26.5 \ \beta \ P^{-0.5}$ above 1 GV.  The agreement is to within $\pm3\%$ up to ~60 GeV/nuc at which point the dependence flattens to $P^{-0.3}$.

**Figure 6:**  Ratio of the individual differential intensity measurements to those at 100 MeV/nuc for H and He (black), and C, (Ne+Mg+Si) and Fe nuclei (red).  This is basically the data in Figure 1 but with the intensities for each species normalized at 100 MeV/nuc.

**Figure 7:** The data in Figure 6 for C, (Ne+Mg+Si) and Fe nuclei intensities (in black) along with the propagation calculations for these nuclei in the LBM for a value of $P_0 = 1.0$ GV (in red).

**Figure 8:** The data in Figure 6 for H and He nuclei intensities (in black) along with the propagation calculations for these nuclei in the LBM for a value of $P_0 = 1.0$ GV (in red).

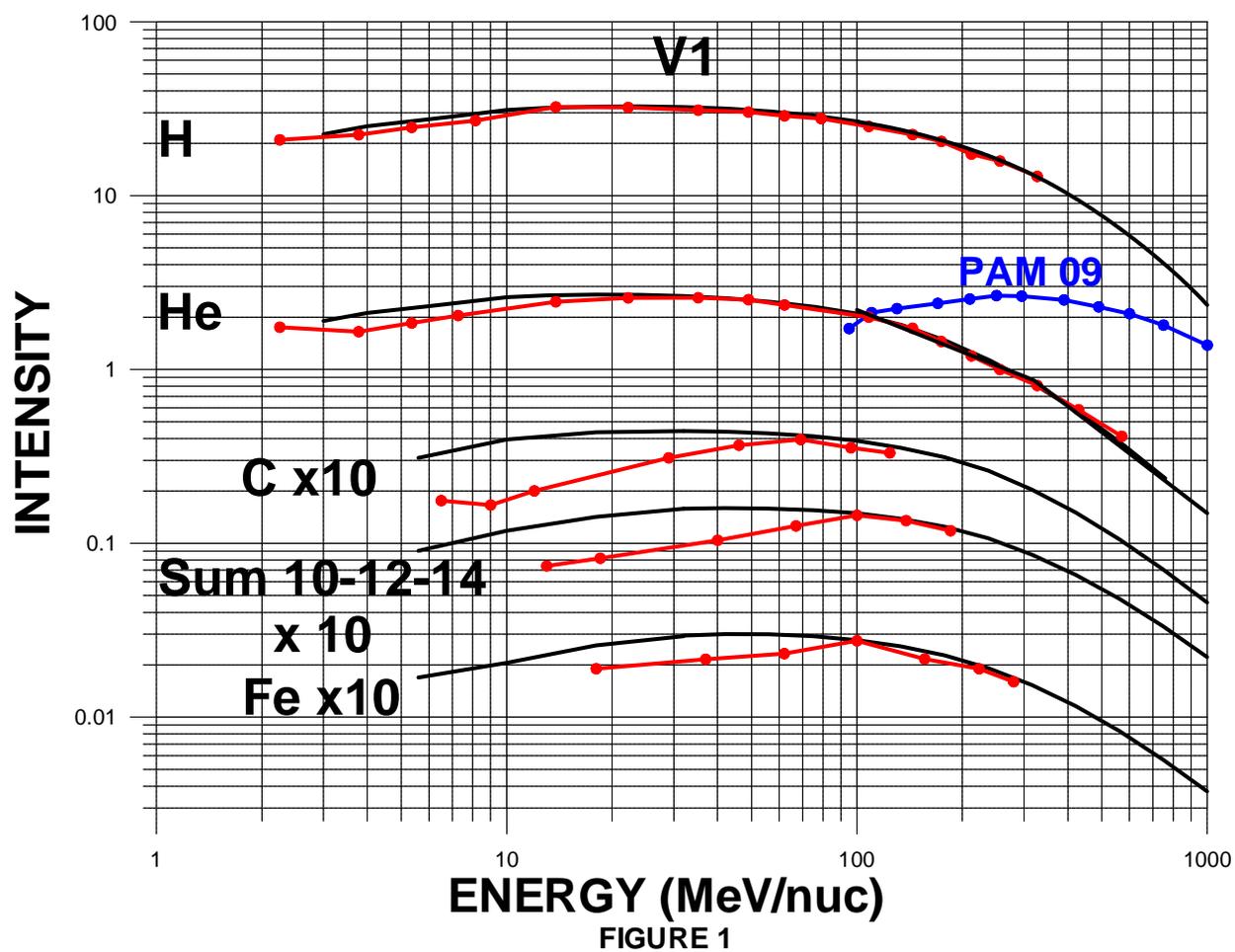

**FIGURE 1**

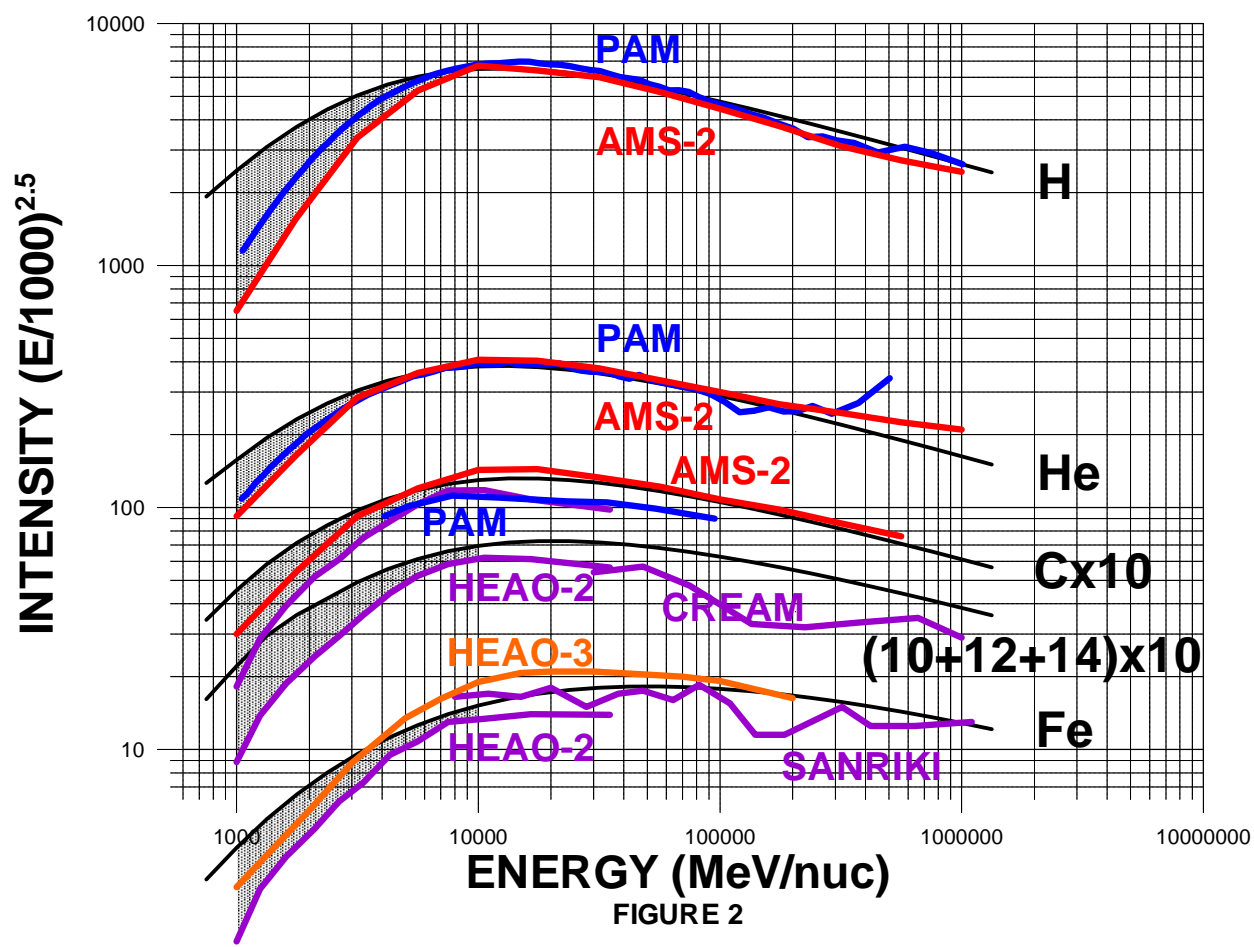

FIGURE 2

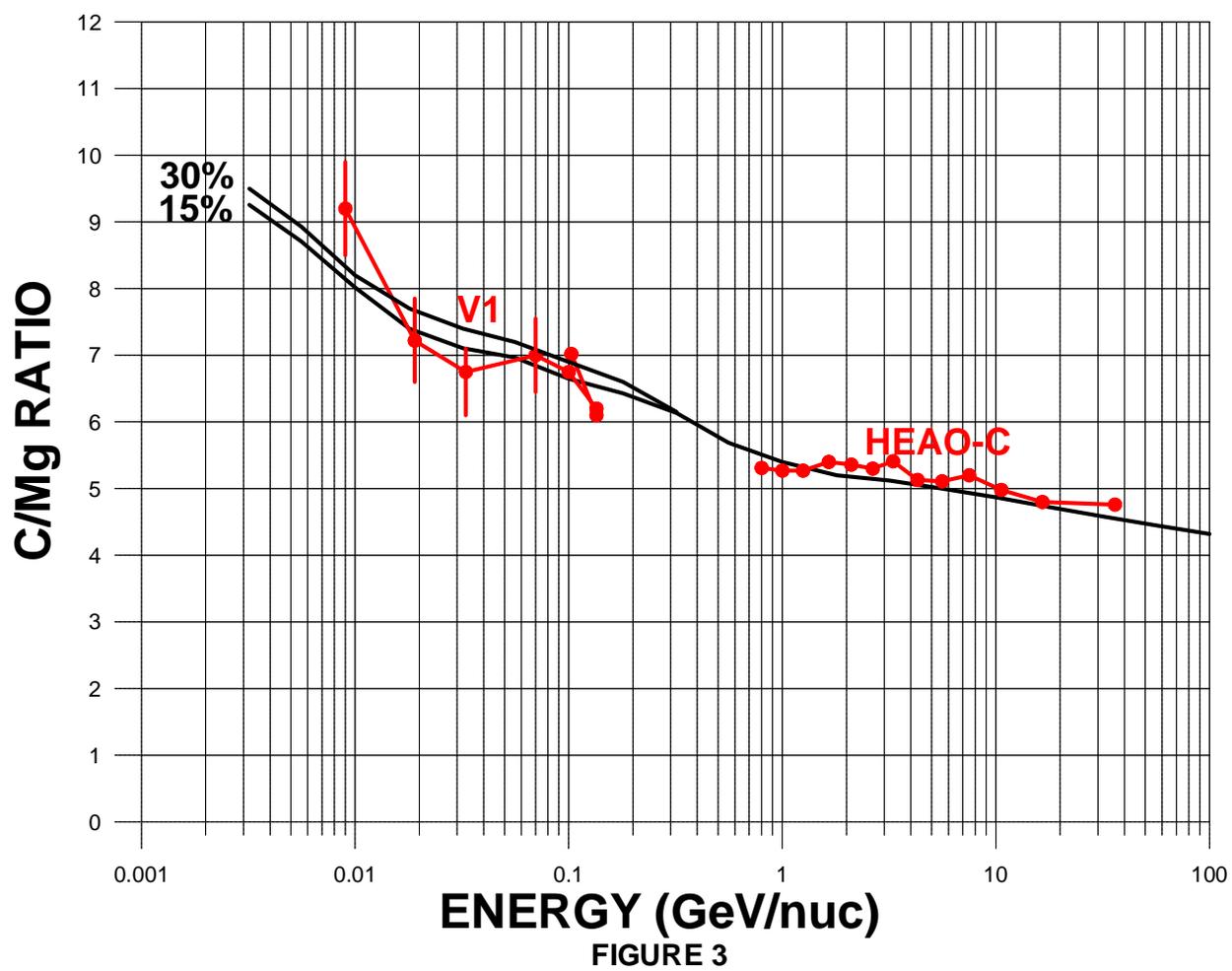

**FIGURE 3**

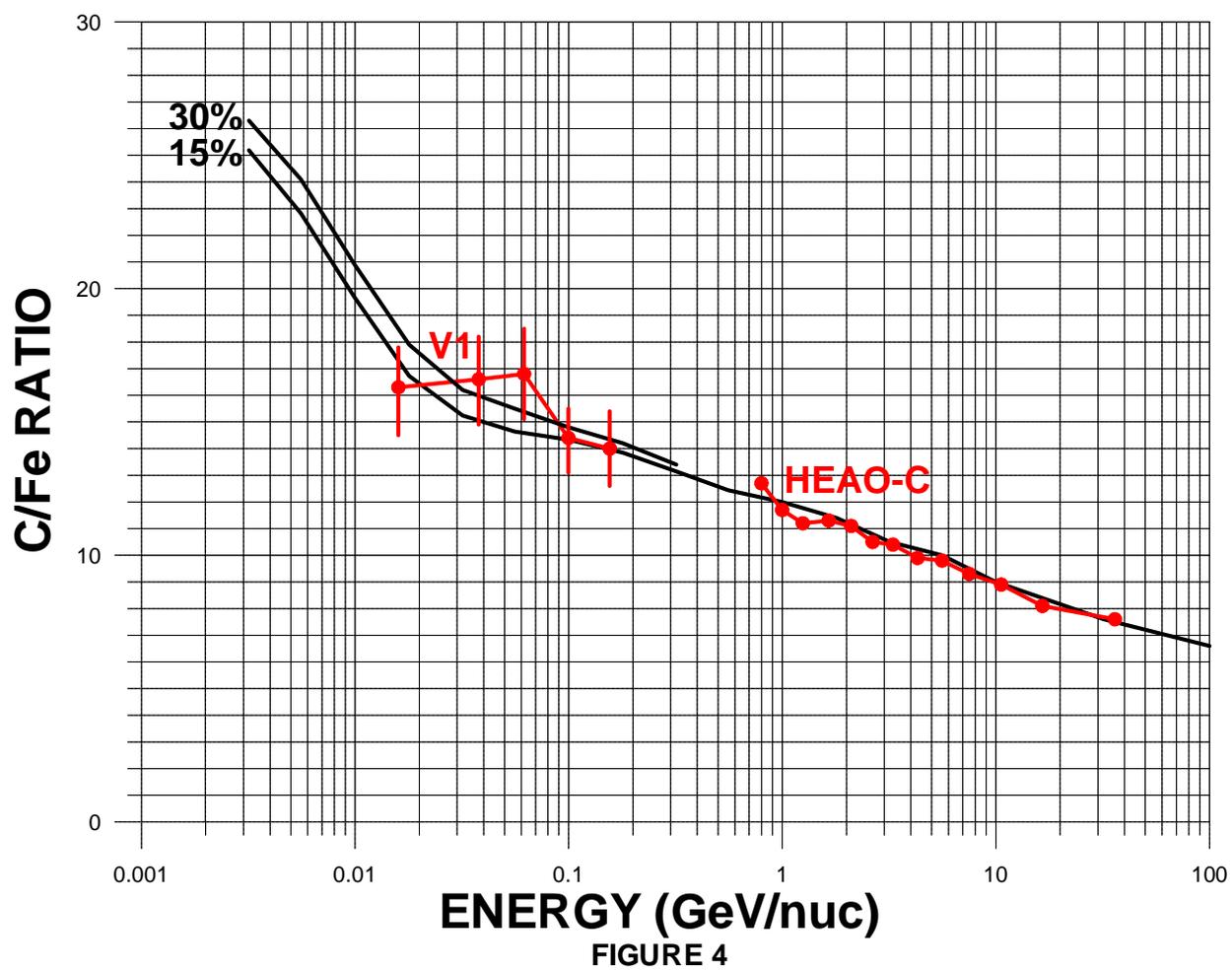

**FIGURE 4**

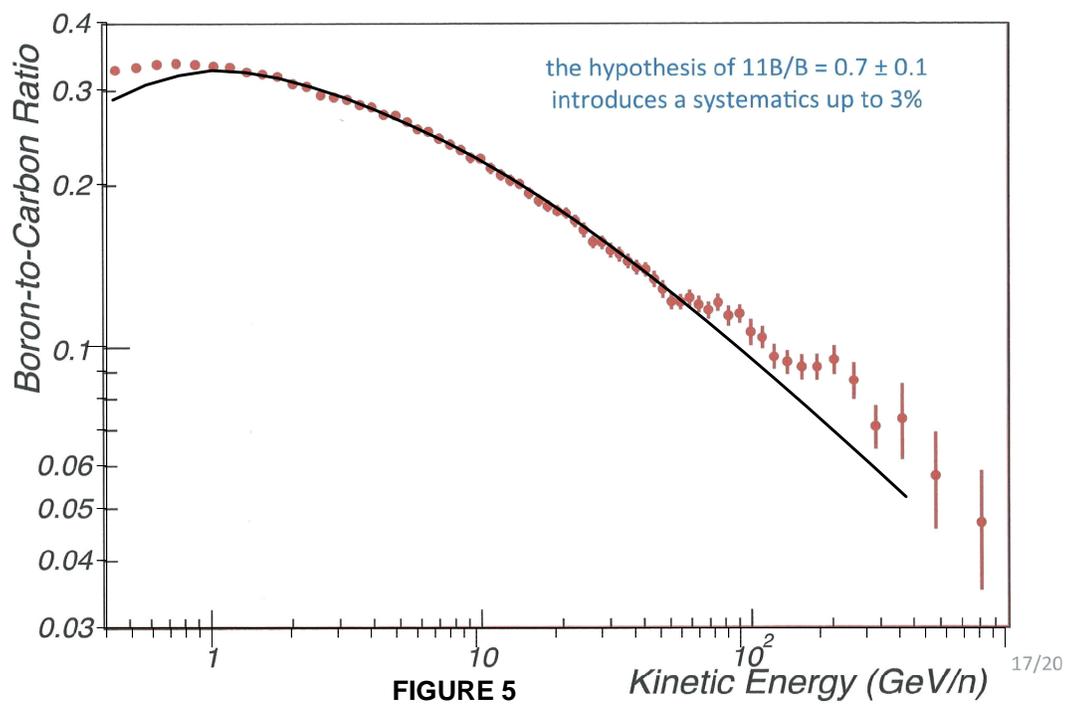

**B/C Ratio converted in Kinetic Energy**

the hypothesis of 11B/B = 0.7 ± 0.1
introduces a systematics up to 3%

**FIGURE 5**

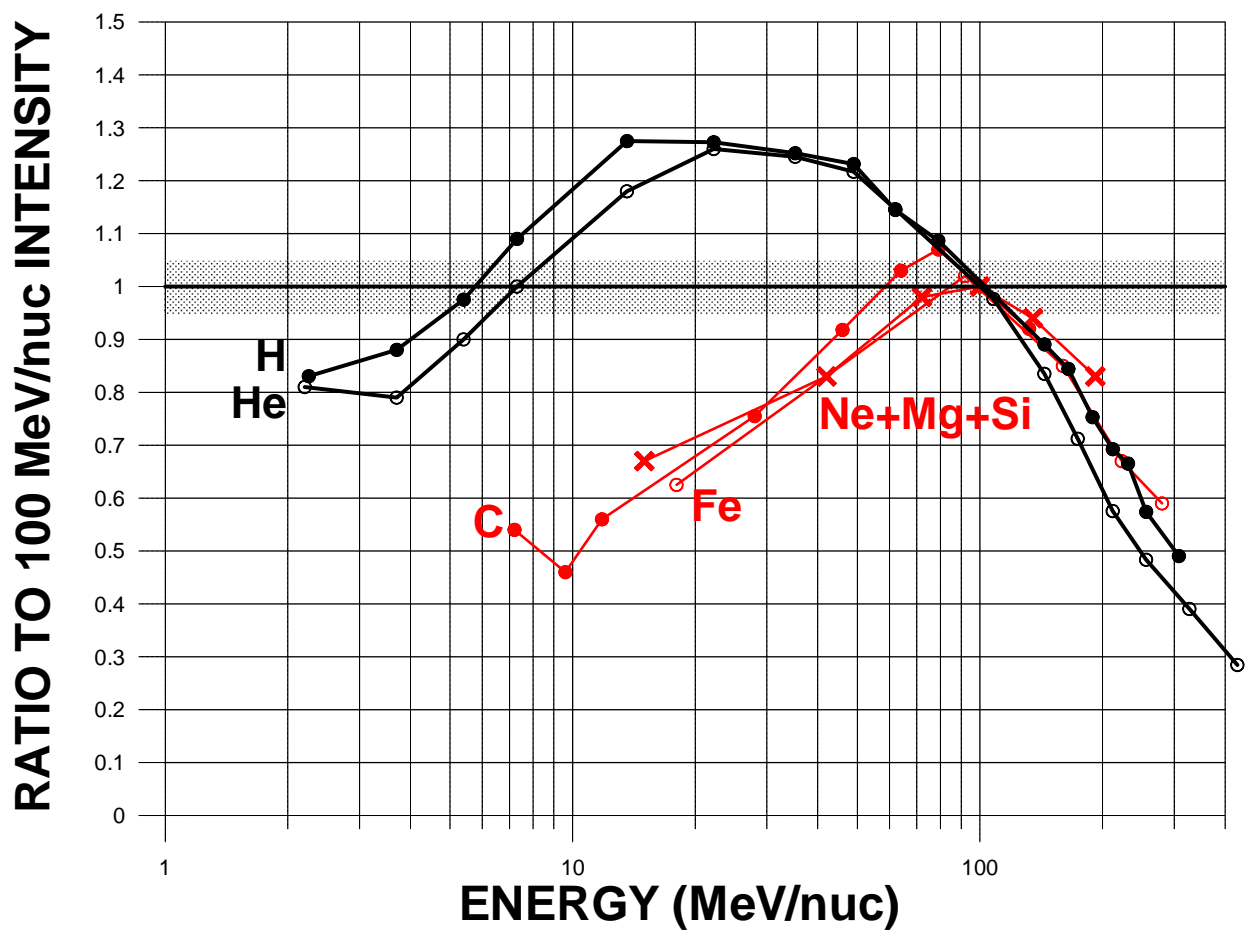

FIGURE 6

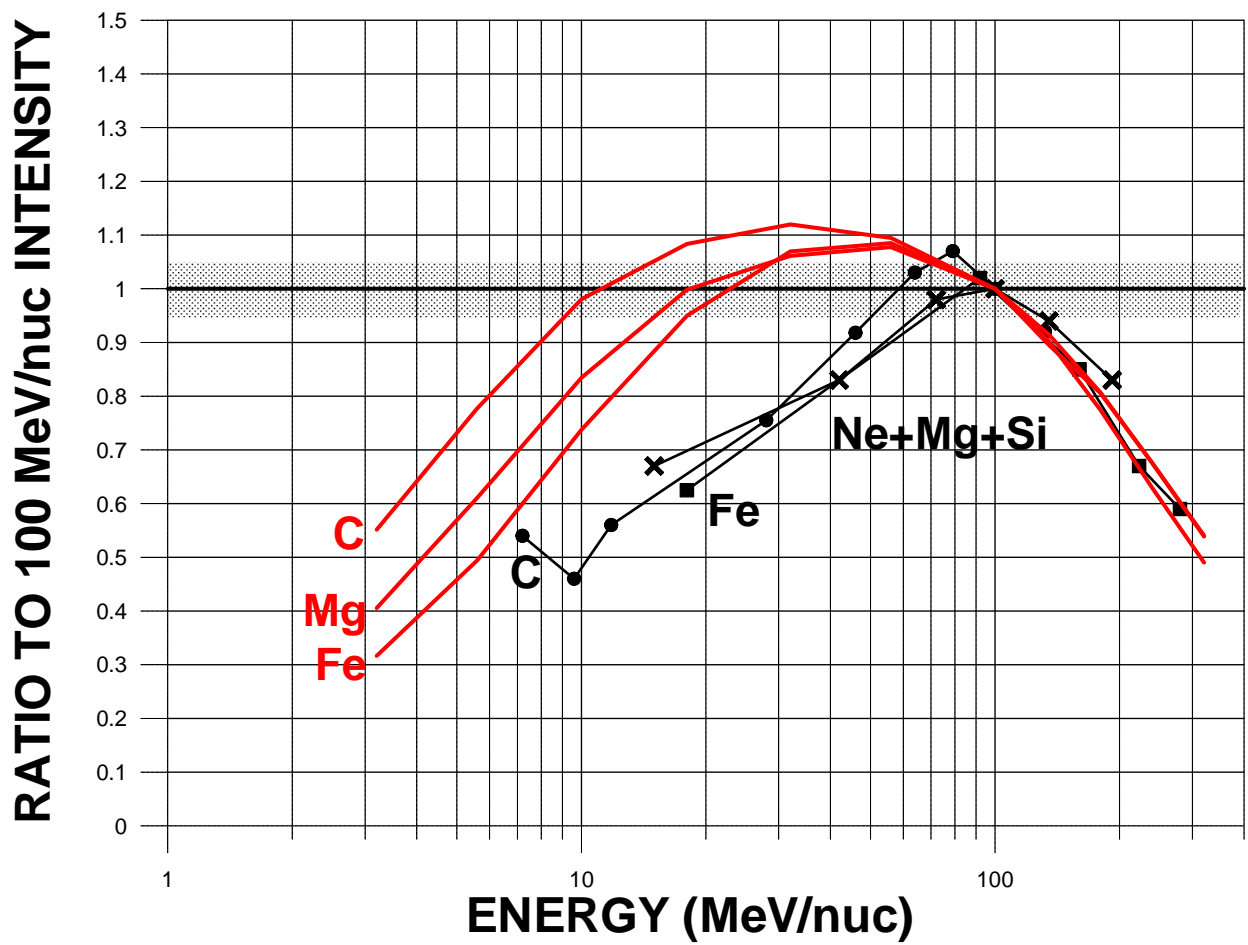

FIGURE 7

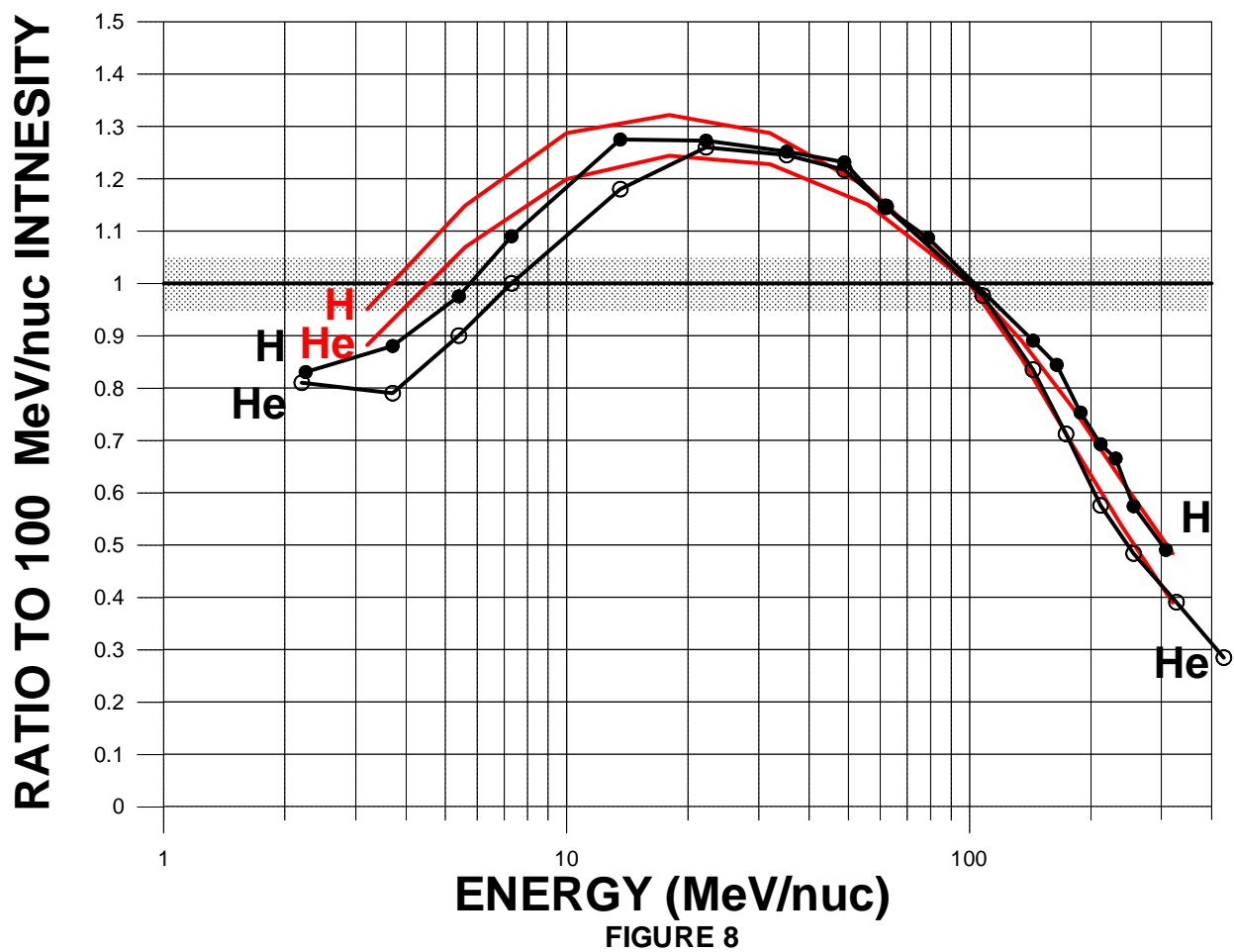

**FIGURE 8**